\documentclass[prd,showpacs,superscriptaddress,nofootinbib,floatfix,preprint,11pt]{revtex4}
\usepackage{graphicx}
\usepackage{amsmath}
\usepackage{amsfonts}    
\usepackage{mathrsfs}
\usepackage{array}
\usepackage[T1]{fontenc}
\usepackage[utf8]{inputenc}
\usepackage{lmodern}
\usepackage[english]{babel}
\usepackage[%
    pdftitle={Affine quantization of Black Holes: thermodynamics, singularity removal and displaced horizons},%
    pdfauthor={Simone Zonetti},%
    pdfkeywords={Black hole thermodynamics, Euclidean path integral, two-dimensional dilaton gravity, affine quantization, spherically symmetric black holes, BTZ black hole, black hole singularities, quantum gravity},bookmarks
]{hyperref}
\newcommand{\mc}[1]{\mathcal{#1}}

\renewcommand{\exp}[1]{\mbox{exp} \left [ #1 \right ]}
\newcommand{\rmi}{\text{\rm{i}}}
\newcolumntype{C}[1]{>{\centering\let\newline\\\arraybackslash\hspace{0pt}}m{#1}}
\begin{document}
\title{Affine quantization of Black Holes: thermodynamics, singularity removal and displaced horizons}

\author{Simone Zonetti} 
\affiliation{Capstone Institute for Theoretical Research, F\'alkag\"otu 17, IS-107 Reykjavik}
\affiliation{Icelandic Meteorological Office, B\'usta\dh avegi 9, IS-105 Reykjavik}
\email{sz@capstone-itr.eu}


\preprint{CITR-14-01}

\date{\today}

\begin{abstract}
In this work we investigate the thermodynamical properties and modified dynamics of black hole solutions in the semiclassical regime of affine coherent state quantization. Using the weak correspondence principle we build a semiclassical effective action that we use to study thermodynamics in the canonical ensemble and to compare classical analytical solutions with the corresponding semiclassical ones.\\
We determine a strong compatibility of these affine black holes with the thermodynamical properties of the classical counterparts, with only minor restrictions on the free parameters of the models. Furthermore we obtain that black hole singularities are removed by quantum effects, resulting also in a modified extension of the horizon, in agreement with the sign of entropy corrections.
\end{abstract}

\maketitle
\tableofcontents
\section{Introduction}
\label{s:Intro}
Rightfully taking their place in popular culture as symbol of mystery, black holes (BH) are one of the best conceptual laboratories for quantum gravity. Unresolved questions in BH physics have attracted efforts in virtually all approaches to quantum gravity, providing also a sort of universal wish list for any viable theory. Alongside the Bekenstein-Hawking entropy/area relation, the removal of BH singularities is perhaps one of the features most sought: it has been obtained within a number of approaches and in a number of different fashions \cite{Bojowald:2007ky,Bojowald:2012qq,Castro:2011fm,Husain:2004yz,Modesto:2004xx,Rovelli:2013osa,Kunstatter:2008qx}, suggesting that some level of convergence is present in the variegated world of quantum gravity theories.\\
In this work we aim at investigating BH solutions in the framework of affine quantization. This approach was first investigated in the late 60s \cite{Aslaksen:1968:URA,Klauder:1967:WCP,Klauder:1965zz} and it has been recently proposed as a possible alternative in the quantization of gravity \cite{Klauder:2011ue,Klauder:2012ij,Klauder:2012vh,Klauder:2001bp,Klauder:1999ba,Watson:2000zp}, as it naturally enforces at the quantum level properties of metric like variables. In particular classical singularities, associated with positive definite variables approaching zero, are removed systematically by (affine) quantum effects.
For instance in \cite{Aslaksen:1968:URA}, for a toy model of gravity, the classical singularities were regularized and in \cite{Klauder:2011ue} it was shown that the affine quantization of the one-dimensional hydrogen atom generates a potential barrier at the scale of the Bohr radius, thus eliminating the Coulomb singularity.\\
In a recent paper \cite{Fanuel:2012fk} we have proved that in a semiclassical limit affine quantization is able to eliminate the initial cosmological singularity in FLRW cosmology, leading to solutions in which the scale factor never vanishes, as a consequence of the implementation of the $a>0$ condition at the quantum level.\smallbreak
In the light of these results it is natural to look into applying the affine quantization scheme to BH solutions, in particular to those simplified models that can provide us with fundamental insight. In order to apply a procedure similar to what developed in \cite{Fanuel:2012fk}, our choice falls on two-dimensional dilaton gravity \cite{Grumiller:2002nm}, also known as 2d Generalized Dilaton Theories (2dGDT): while providing us with a simplified framework, they allow for a comprehensive description of a large number of BH models, with an extensive literature on BH solutions and their properties \cite{Bose:1995gy,Cruz:1996zy,Grumiller:2004wi,Katanaev:1996ni,Kummer:1996hy,LouisMartinez:1993eh,Grumiller:2006rc}. In particular all classical solutions are known \cite{Grumiller:2002nm} and a thorough description of BH thermodynamics is presented in \cite{Grumiller:2007ju}.\\
In the following we will apply affine coherent state quantization to effective BH models derived from 2dGDT, which we will then simply call \emph{affine black holes} (ABH), in order to obtain a semiclassical model including quantum corrections. We will then proceed in the analysis of BH solutions, discussing their thermodynamical properties and comparing their behaviour in the classical and semiclassical regimes.\\
This paper is organized as follows: in Section \ref{s:effective} we will develop an effective model for BH starting from the full 2dGDT action, while in Section \ref{s:affine} we will discuss the generalities of affine coherent state quantization and calculate quantum corrections for our specific case.\\
Section \ref{s:physics} first reviews the canonical ensemble approach to BH thermodynamics and then deals with its adaptation to ABH. The general procedure for the comparison of classical and semiclassical (numerical) solutions is also outlined. Section \ref{s:models} discusses thermodynamical properties, singularity removal and horizon corrections for a number of dilaton gravity inspired ABH models. Finally in Section \ref{s:discussion} we summarize the results and briefly outline some possible developments.
\section{Effective one-dimensional model for black hole solutions}
\label{s:effective}
Let us start by considering the general two-dimensional dilaton gravity action \cite{Grumiller:2002nm}:
\begin{equation}\label{eq:DGaction}
S_{dg}=-\frac{1}{2}\int dx^{2}\sqrt{-g}\left(XR-U(X)\left(\partial X\right)^{2}-2V(X)\right)\ ,
\end{equation}
The dilaton $X$ is defined by its coupling to the two-dimensional Ricci scalar $R$ and different models are distinguished by different choices of the potentials $U,V$ (see \cite{Grumiller:2006rc} for a summary). Equations of motion are:
\begin{subequations}\label{eq:DGeom}
\begin{align}
\delta g^{\mu\nu}:& -\frac{1}{2}g_{\mu\nu}U(X)\left(\partial X\right)^{2}-g_{\mu\nu}V(X)-U(X)\partial_{\mu}X\partial_{\nu}X+\partial_{\mu}\partial_{\nu}X+\partial^{2}Xg_{\mu\nu}=0\ ,\\
\delta X:& R+U'(X)\left(\partial X\right)^{2}-2V'(X)+2U\partial^{2}X=0\ ,
\end{align}
\end{subequations}
where the prime denotes derivatives with respect to $X$. Solutions to the associated equations always possess a Killing vector $\partial_t$, with orbits where $X=const$. Using a diagonal gauge:
\begin{subequations}\label{eq:diaggauge}
\begin{align}
ds^{2}=&-\xi(r)dt^{2}+\frac{1}{\xi(r)}dr^{2} \ ,\\
X=&X(r) \ ,
\end{align}
\end{subequations}
we have solutions of the form:
\begin{subequations}\label{eq:DGsol}\begin{align}
X_r=&e^{-Q(X)} \ , \label{eq:Xsol} \\
\xi=&e^{Q(X)} \left( w(X)-2M \right) \ . \label{eq:XIsol}
\end{align}\end{subequations}
Notice that the subscript $r$ denotes derivation, \emph{i.e.} $X_r=\partial_r X=X_{,r}$ and $r\in\mathbb{R}$. The $Q$ and $w$ functions are defined in terms of the original potentials $U,V$ as:
\begin{subequations}\label{eq:QWUV}\begin{align}
Q(X)&=Q_{0}+\int^{X}d\tilde{X}U(\tilde{X)}\ ,\\
w(X)&=w_{0}-2\int^{X}d\tilde{X}V(\tilde{X})e^{Q(\tilde{X})}\ .
\end{align}\end{subequations}
While three constants are present, $Q_0$, $w_0$ and $M$, $w_0+M$ represents a single parameter that can be chosen in a way that we can restrict ourselves to $M\ge0$ for physical solutions. It is clear that a zero in $\xi(X)$ corresponds to a vanishing Killing norm $\sqrt{\partial_t}$, hence to a Killing horizon. We can therefore consider as BH all physical solutions with $M>0$ that exhibit horizons and focus on this class of solutions in the rest of this paper. For further details, please refer to \cite{Grumiller:2007ju}.\\
Let us now look at the two dynamical variables $X$ and $\xi$. While horizons are found at $X=X_h$, so that $w(X_{h})=2M$ and $\xi(X_h)=0$, curvature singularities, when present, are associated with $X\to0$, when $U(X)$ is singular itself (see \cite{Grumiller:2006rc}). Therefore physical non-singular solutions would naturally carry the condition $X>0$. This positivity condition, in the same way as the $a>0$ condition in FLRW cosmology, serves very well in the affine quantization scheme. On the other hand the metric function $\xi(r)=\xi(X)$ sees no restrictions to its sign and determines the classical geometry of space-time in terms of the dilaton $X$. In particular it is by enforcing \eqref{eq:XIsol} at the classical level that we look specifically at BH solutions.\\
If our purpose is indeed to compare classical and semi-classical BH, it is reasonable to look at $\xi=\xi(X)$ as a purely classical quantity that constitutes a background to our model and leave it out of the affine quantization procedure, introducing quantum corrections only through the dynamics of $X$. We are then looking at a classical geometry $\xi$ induced by the dynamics of a quantum field $X$. While this is an arbitrary choice - we could have chosen the dilaton to be classical and $\xi$ to be quantized - we are interested in looking at how the physical condition $X>0$ can be implemented at the quantum level and how this affects know classical BH.\\
To this end we can interpret our two-dimensional (static) BH solutions in terms of an $r$-reparametrization invariant effective model consisting of a single physical degree of freedom $X(r)$ whose dynamics is governed by the equations of motion \eqref{eq:DGeom}, with the gauge conditions \eqref{eq:diaggauge} and the relation \eqref{eq:XIsol}. After a simple substitution we are left with two independent equations:
\begin{subequations}\label{eq:effeq}
\begin{align}
X_{rr}+e^{-2Q}Q'=0\ ,\\
X_{r}^{2}-e^{-2Q}=0\ ,
\end{align}
\end{subequations}
which we can see as an equation of motion proper and a constraint, as we would expect from the requirement of $r$-reparametrization invariance. If we take a Lagrangian of the form:
\begin{equation}\label{eq:Leff}
\mc{L}_{eff}= N(r)^{-1}g(r)X_{r}^{2}+N(r)e^{-2Q(X)} \ ,
\end{equation}
where $N(r)$ plays the role of a Lagrange multiplier and $g(r)$ is a background metric, it is straightforward to see that \eqref{eq:effeq} can be generated when we impose the gauge $N=g=1$. This Lagrangian is diff-invariant if we consider the transformation laws $X_{r}\to\epsilon_{r}X_{r}$, $g\to\epsilon_{r}^{-2}g$ when $r\to \epsilon(r)$.\\
We can therefore look at \eqref{eq:Leff}, supplied by \eqref{eq:XIsol}, as an effective model that yields the same physical solutions (with identical mass) as full 2dGDT \eqref{eq:DGaction} with the gauge choice \eqref{eq:diaggauge}.\\
The corresponding Hamiltonian to \eqref{eq:Leff}, starting point for affine quantization, is:
\begin{equation}\label{eq:Heff}
\mc{H}_{eff}=N\left(\frac{P_{X}^{2}}{4g}-e^{-2Q}\right) \ ,
\end{equation}
where the momentum for $X$ is $P_{X}=\frac{2g}{N}X_{r}$ and one has the constraints $P_N=0$ and $\mc{H}_{eff}=0$, as expected.
\section{Affine coherent state quantization}
\label{s:affine}
While phase space is build with the canonical algebra $\{X,P_X\}=1$, the physical requirement $X>0$ gives difficulties in the definition of self-adjoint operators $\hat{X}, \hat{P}$ with canonical commutation relations $[\hat{X},\hat{P}_X]=\rmi\hbar$. \cite{Klauder:1970ut}. A possible solution to this issue is affine coherent state quantization, built around the affine variables $X,D$, with $D=XP_X$, where the positivity condition $X>0$ is naturally implemented for quantum operators, which can be made self-adjoint. Let us review the quantization procedure, referring to \cite{Klauder:2012uf} and references therein for further details.
\subsection{The affine algebra}
Let us consider the canonical classical variables $\{q,p\}=1$, with the requirement $q>0$. The corresponding affine algebra will be $\{q,d\}=q$, with $d=qp$. The main reason to pick $d$ over $p$ is that $d$ acts dilating $q$, thus preserving a lower bound $q>0$, while $p$ generates translations. Introducing the operators $Q,P$, satisfying the commutation relation $[Q,P]=\rmi\hbar$, we obtain, multiplying both sides by $Q$:
\begin{equation}
[Q,D]= \rmi \hbar Q \qquad D = \frac{1}{2}\left( P Q + Q P \right) \ .
\end{equation}
This takes the name of \emph{affine commutation relation}, as it refers to the affine group of transformations $q\to q'=a q + b$. To represent the operators on $q-$space we can take:
\begin{subequations} \label{eq:DQ_rep}
\begin{align}
D f(q)& =-\rmi\hbar \ q^{1/2} \partial_{q}(q^{1/2}f(q))\label{eq:D_rep}\ ,\\
Q f(q)& = q f(q) \ ,
\end{align}
\end{subequations}
which clearly shows the dilating effect of the $D$ operator. In order to ensure that the operator $D$ is self-adjoint, we will have to require that the boundary term coming from:
\begin{equation} \label{eq:selfadjcond}
\langle \phi|D\psi\rangle-\langle D^{\dagger} \phi|\psi\rangle=-\rmi\hbar \int_{0}^{+\infty} d q \ \partial_{q}[\phi^{*}(q)\ q\ \psi(q)]
\end{equation}
vanishes, for square integrable functions $\psi(x)\in \text{Dom }  D$ and $\phi(x)\in \text{Dom } D^{\dagger}$ defined on the half-line $q>0$. In particular these functions need to satisfy:
\begin{equation}
\lim_{q\to 0}x^{1/2}\psi(x)=0=\lim_{x\to 0}x^{1/2}\phi(x) \ ,
\end{equation}
conditions ensuring that \eqref{eq:selfadjcond} vanishes, so that the domains of $D$ and $D^\dagger$ coincide. The self-adjoint operators $Q,D$ will obey the algebra $\left[Q,D\right]=\rmi \hbar Q$, which guarantees the existence of a unitary irreducible representation of operators with a positive spectrum for $Q$ \cite{Aslaksen:1968:URA}.
\subsection{Affine coherent states}
Affine coherent states are defined by:
\begin{equation}
|p,q\rangle=e^{\rmi p Q/\hbar}e^{-\rmi \ln(q/\mu)D/\hbar}|\eta\rangle\label{eq:affinecoherent}\ ,
\end{equation}
defined on $\mathbb{R} \times \mathbb{R}^+$, where $\mu$ is a scale with dimension of length and $|\eta\rangle$ is some fiducial unit vector of choice. In particular we can pick it to satisfy the polarization condition 
\begin{equation} \label{eq:polarizationcondition}
\left[\frac{Q}{\mu}-\mathbb{I}+\rmi \frac	{D}{\beta\hbar} \right]|\eta\rangle=0\ ,
\end{equation}
with $\beta$ a free dimensionless parameter. The wave function for the fiducial vector can be easily calculated and takes the form:
\begin{equation} \label{eq:fiducialwavef}
\langle q|\eta\rangle=N_\eta (q/\mu)^{\beta-1/2}\exp{(-\beta q/\mu)}\ ,
\end{equation}
with the normalization factor $N_\eta=(2^{-2\beta}\beta^{-2\beta}\Gamma[2\beta]\mu)^{-1/2}$. This implies also $\langle\eta|Q|\eta\rangle=\mu$ and $\langle\eta|D|\eta\rangle=0$. The functional representation of affine coherent states is given by:
\begin{equation}\label{eq:affinecoherentwf}
\langle x|p,q\rangle = N e^{\rmi p x/\hbar}\left(\frac{x}{q}\right)^{\beta}x^{-\frac{1}{2}}\exp{-\frac{\beta x}{q}}\ ,
\end{equation}
with $N=(2^{-2\beta}\beta^{-2\beta}\Gamma[2\beta])^{-1/2}$. Notice that, remarkably, no dependence from the scale $\mu$ is present. The average values for $Q,D$ are now readily calculated, giving:
\begin{subequations} \label{eq:QDaveragevalues}
\begin{align}
\langle p,q|Q|p,q\rangle&=q \ , \\
\langle p,q|D|p,q\rangle&=p\ q \ ,
\end{align}
\end{subequations}
as we would expect. The wave function \eqref{eq:affinecoherentwf} can be used to show that:
\begin{equation}
\mathbb{I}=\int \! \! \frac{d p\ d q}{2\pi\hbar} \ \frac{|p,q\rangle\langle p,q|}{\mu \langle\eta|Q^{-1}|\eta\rangle} \ ,
\end{equation}
is indeed a resolution of identity, for instance reproducing $\langle x' | x \rangle = \delta(x-x')$, with the constraint $\beta>1/2$, which ensure that we avoid the simple poles of the $\Gamma$ function located at negative integer values of the argument.\\
Let us take a closer look at the parameters $\mu$ and $\beta$ which appear in this construction by means of a comparison with \emph{canonical} coherent states. The analogue of \eqref{eq:fiducialwavef} for canonical coherent states takes the form:
\begin{equation}
\langle q| \Omega\rangle =\left(\frac{\pi\hbar}{\lambda_{0}}\right )^{-1/4}e^{-\lambda_{0}q^{2}/2\hbar}\ ,
\end{equation}
with a fiducial vector $|\Omega\rangle$ satisfying the polarization condition
\begin{equation}
\left[ \frac{P}{\lambda_{p}}-\rmi \frac{Q}{\lambda_{q}}\right]|\Omega\rangle=0
\end{equation} and $\lambda_0 = \lambda_p/\lambda_q$. In this case $\lambda_q$ (resp. $\lambda_p$) is setting the width of Gaussian in the functional representation of the fiducial vector in terms of $q$ (resp. $p$), and one usually chooses the two scales in a way that $\lambda_0=1$. Similarly, $\mu$ controls the width of the wave function \eqref{eq:fiducialwavef} and the average value of $Q$. Furthermore, by comparing the two polarization conditions, $\beta$, or equivalently $\beta \hbar$, plays a role similar to the dimensionless product $\lambda_q \lambda_p$, which contains information about the relative weight of $P$ and $Q$ in the construction of canonical coherent states. In view of this interpretation, $\beta$ can be seen as a free parameter labelling different representations of the same physical states and can therefore be chosen to our convenience.
\subsection{The Weak Correspondence Principle}
Let us briefly review the basic definitions and statement of the Weak Correspondence Principle, as introduced by Klauder in \cite{Klauder:1967:WCP}: generally speaking we are looking for a correspondence rule between a classical system and its quantum counterpart that, through the use of (affine) coherent states, incorporates quantum dynamics in a classical Hamiltonian theory.\\
We consider a set of conjugate operators $Q,P$, with canonical commutation relations $[Q,P]=\rmi \hbar$, and we build a family of unitary Weyl operators (cf. \eqref{eq:affinecoherent}):
\begin{equation}
U[p,q]\equiv \exp{\rmi \left( pQ-Pq \right)} \ ,
\end{equation}
determining a set of unit vectors:
\begin{equation}
|p,q\rangle \equiv U[p,q] |\eta \rangle\ ,
\end{equation}
where $|\eta\rangle$ is a unique state satisfying a specific polarization condition (cf. \eqref{eq:polarizationcondition}). Generally speaking the states $|p,q\rangle$, referred to as the \emph{overcomplete family of states} (OFS), are not mutually orthogonal but, on the contrary, the overlap between different states contains information about the system and commutation relations. In addition, while $|p,q\rangle$ do not necessarily span the full Hilbert space, the Weak Correspondence Principle is valid in whichever subspace the OFS spans. Our interest is to consider the diagonal OFS matrix elements:
\begin{equation}\label{eq:WCP}
h(p,q)=\langle p,q| \mathcal{H}|p,q \rangle 
\end{equation}
for a quantum Hamiltonian operator $\mathcal{H}$, to be related to the classical Hamiltonian $H(p_c,q_c)$, with classical canonical variables $\{q_c,p_c\}=1$. This relation is usually taken to be:
\begin{equation}\label{eq:quantumop}
\mathcal{H}[P,Q]=:H(P,Q): \ ,
\end{equation}
where $:$ denotes normal ordering. Using the unitary operator $U$, which generate translation in both $P$ and $Q$, we can see that in fact:
\begin{equation}
\langle \eta | U[p,q]^\dagger :H(P,Q): U[p,q] | \eta \rangle = \langle \eta | :H(P+q,Q+p): |\eta\rangle = h(p,q)\ .
\end{equation}
Therefore a direct consequence of \eqref{eq:quantumop} is that diagonal OFS matrix elements of a quantum operator yield to the corresponding classical functional with the identification:
\begin{equation}
p=p_c\qquad q=q_c\ ,
\end{equation}
for all consistent choices of normal ordering. A fundamental feature of this construction is that the relation \eqref{eq:WCP} is valid disregarding whether \eqref{eq:quantumop} holds or not, for instance in the case in which there is no irreducible representation for $P,Q$ and therefore $\mathcal{H}$ is not only a function of $P,Q$.\\
Moreover, if the overlap function of states $\langle p',q' | p,q\rangle$ never vanishes, any composite (polynomial) operator in the domain of the OFS can be represented in terms of its diagonal matrix elements as\cite{Klauder:1968dq,Klauder:2012uf}:
\begin{equation}
\mathcal{A}=\int \frac{dp\ dq}{2\pi \hbar}a(p,q) |p,q\rangle \langle p,q| \ .
\end{equation}
\subsection{The extended Hamiltonian for affine coherent states}
Let us look once again at a generic canonical quantum theory described by an Hamiltonian $\mathcal{H}(P,Q)$. The associated Schr\"ordinger equation, $\rmi \hbar \partial_t |\psi(t) \rangle = \mathcal{H} |\psi(t)\rangle$, can be formally obtained from a quantum action functional:
\begin{equation} \label{eq:qaction}
S_Q=\int^T_0 \! dt\ \langle \psi(t) | \left( \rmi \hbar \partial_t - \mathcal{H} \right) |\psi(t)\rangle
\end{equation}
by variation w.r.t. unit vectors $\langle \psi |$ and $|\psi\rangle$, considered as independent variables. In order to look at a regime in which classical and quantum dynamics coexist, rather than allowing an arbitrary set of quantum states to enter the quantum action functional, we can adopt the view point of the Weak Correspondence Principle and we restrict ourselves to those states that are accessible to a macroscopic observer, namely coherent states: canonical coherent states are those that translate a system or put it in motion with constant velocity (since $\dot{q}=\partial H_c / \partial p$), while for affine coherent states translation is replaced by (de)magnification.\\
By limiting the set of quantum states $|\psi\rangle$ to to affine coherent states $| p(t),q(t) \rangle$ we can explicitly calculate \eqref{eq:qaction}, considering that, with the fiducial vector defined by \eqref{eq:fiducialwavef}, the overlap function reads:
\begin{equation}
\langle p',q'|p,q\rangle=2^{2\beta}\left(qq'\right)^{\beta}\left((q+q')-\frac{i(p-p')qq'}{\hbar}\right)^{-2\beta}\ ,
\end{equation}
which is non-vanishing on all $\mathbb{R}\times\mathbb{R}^+$.\\
For the first term, using \eqref{eq:D_rep} on \eqref{eq:affinecoherentwf} we have:
\begin{equation}
\langle p(t), q(t) | \rmi \hbar \partial_t |p(t), q(t) \rangle = \rmi \hbar \int_0^\infty dx || \langle x|p,q\rangle ||^2 \left(\frac{ \rmi \dot{p}x}{\hbar}+\beta\frac{\dot{q}}{q}\left(\frac{x}{q}-1\right)\right) = -\dot{p}q \ ,
\end{equation}
while it is possible to check that for any composite operator $\mathcal{H}$ we have the identity:
\begin{equation}
h(p,q):=\langle p,q|\mathcal{H}(D,Q)|p,q\rangle=\langle\eta|\mathcal{H}\left(D+\frac{pq}{\mu}Q,\frac{q}{\mu}Q\right)|\eta\rangle \ ,
\end{equation}
that operatively defines the Extended Hamiltonian $h(p,q)$, with ordering chosen in a way that $Q$ and $D$ operators alternate, for instance $:Q^{-3}D^2:=Q^{-1}DQ^{-1}DQ^{-1}$, consistently with the so called ``anti-Wick quantization'' rule. At this point we can use the restricted quantum action principle:
\begin{equation}
S_{Q(R)} =\int^T_0 \! dt\ \left( p\dot{q} - h(p,q) \right)
\end{equation}
to determine the equations of motion as we would do with any classical canonical system. Let us stress that the quantum dynamics contained in $h$ is not coming from a series expansion in powers of $\hbar$, but rather from a restriction to coherent states: this has the advantage of allowing us to look at regimes in which quantum contributions are expected to be large, e.g. close to gravitational singularities, without contradicting our own construction.\\
Before going back to our effective BH model, let us point out some general features for the extended Hamiltonian. Let us look at a rather general classical Hamiltonian in the form:
\begin{equation}\label{eq:classheg}
\mc{H}(p,q)=A q^{-n+2} p^2+B q^{-m+1} p-\sum V_k q^k=A q^{-n} d^2+B q^{-m} d-\sum V_k q^k\ ,
\end{equation}
with constant $A,B$, and a classical restriction to $q>0$. In order to obtain $h(p,q)$ we need to calculate matrix elements:
\begin{subequations} \label{eq:amatrixe}
\begin{align}
\mu^{n}\langle Q^{-n}\rangle&=\delta(n)\ ,\\
\mu^{n}\langle Q^{-n/3}DQ^{-n/3}DQ^{-n/3}\rangle&=\gamma(n)\ ,\\
\mu^{n}\langle Q^{-n/2}DQ^{-n/2}\rangle&=0\ ,\\
\langle Q^{-n/3}DQ^{-2n/3}\rangle&=-\langle Q^{-2n/3}DQ^{-n/3}\rangle\ ,
\end{align}
\end{subequations}
where:
\begin{subequations}\label{eq:deltagamma}
\begin{align}
\delta(n)  = & \left(2\beta\right)^{n}\frac{\Gamma(2\beta-n)}{\Gamma(2\beta)} > 1\ ,\\
\gamma(n)  = & \hbar^{2}\left(2\beta\right)^{n}\frac{18\beta+n(n-9)}{36}\frac{\Gamma(2\beta-n)}{\Gamma(2\beta)} > 0\ ,
\end{align}
\end{subequations}
requiring $\beta>n/2$ in order to be finite and avoid the poles of the $\Gamma$ function at negative integers. Then for $h(p,q)$ we obtain:
\begin{equation}\label{eq:extheg}
h(p,q)=A \delta(n) q^{-n+2} p^2+B \delta(m) q^{-m} p- \sum V_k \delta(-k) q^k + A \gamma(n) q^{-n} \ ,
\end{equation}
where we see quantum corrections to the classical terms (the $\delta$'s) and the appearence of an additional dynamical term generated from the kinetic term itself, enforced by the coefficient $\gamma>0$. This term can determine the avoidance of singularities located at $q\to0$, as one can see from the trivial example with $n=0,A=1,B=0$ and $V_{-1}=1,V_{k\neq-1}=0$. The classical and extended Hamiltonians are:
\begin{equation}
\mc{H}(p,q)=p^{2}-q^{-1} \ ,\qquad h(p,q)=\delta(2)p^{2}-\delta(1)q^{-1}+\gamma(2)q^{-2} \ ,
\end{equation}
so that the classical singularity is avoided by the repulsive potential $\gamma(2)q^{-2}$, which is dominant at $q\sim0$. This effect is clearly limited by the value of the exponent $n$ in \eqref{eq:extheg}, which can cure singularities up to order $n$ in the potential $V$.\\
It is easy to check that in the classical limit \eqref{eq:extheg} reproduces \eqref{eq:classheg} if we consider that:
\begin{equation}
\lim_{\beta\to\infty}\delta(n)=1\qquad\lim_{\beta\to\infty,\beta\hbar\to0}\gamma(n)=0 \quad \Longrightarrow \quad \lim_{\beta\to\infty,\beta\hbar\to0}h(p,q)=H(p,q)\ .
\end{equation}
Let us give a further look at the the coefficients $\delta$ and $\gamma$ in \eqref{eq:deltagamma}: while the latter is a genuine quantum correction, $\hbar$-dependent, the former is a function of $\beta$ alone. We have already mentioned that the role of $\beta$ is to label different representations of physical states; furthermore we have just seen that a lower bound $\beta>n/2$ is introduced by finiteness requirements on the matrix elements \eqref{eq:amatrixe}, with $n$ being the largest negative power of the coordinate variable that appears in the classical Hamiltonian. The bound is therefore model specific: for instance in the case of the \emph{ab}-family models described below we have to allow for terms of the form $q^{-a}$, where $a\in \mathbb{R}$, so that we need $\beta > a / 2$. Schwarzschild-(A)dS is an extreme example, since due to a potential $e^{Q^{-1/2}}$ its extended Hamiltonian is only defined in the $\beta\to\infty$ limit.\\
Then, since $\beta$ has no direct physical interpretation, we can look at the case $\beta \to \infty$ with no loss of generality: considering that the $\delta$'s reduce to unity, while $\gamma$ is strictly positive if we leave $\hbar\neq0$, taking this limit has the double advantage of simplifying the semiclassical picture, leaving $\hbar$-dependent corrections only, and ridding us of the need of choosing $\beta$ in each case. Let us stress that all results in this paper can be obtained for finite $\beta$ with a few additional, but minor, complications. A discussion about the case with finite $\beta$ is contained in appendix.\\
Taking then $\beta \to \infty$ and rescaling $\hbar$ accordingly we can bring the extended Hamiltonian \eqref{eq:extheg} to be just the classical Hamiltonian plus a quantum correction:
\begin{equation}
h(p,q)=A q^{-n+2} p^2+B q^{-m} p- \sum V_k q^k + A \gamma(n) q^{-n} = \mc{H}(p,q)+ A \gamma(n) q^{-n}\ .
\end{equation}
Let us point out that this limit is only meaningful if applied on the semiclassical model and not in the full quantum regime, where the coherent state construction requires a finite value for $\beta$. Let us now go back to our BH models, described by the classical Hamiltonian \eqref{eq:Heff}, and pick the gauge $N=g=1$. With the procedure above we can easily obtain the extended Hamiltonian as:
\begin{equation}\label{eq:Haffinedelta}
h(X,P_X)=\frac{\delta(2)}{4}P_{X}^{2}+\sum Q_{n}X^{n}\delta(-n)+\frac{\gamma(2)}{4}X^{-2} \ ,
\end{equation}
where the exponential potential has been expanded as $e^{Q(X)}=\sum Q_{n}X^{n}$. Finiteness of the coefficients imposes $\beta>1$, to be supplemented by conditions coming from the potential $Q$. By taking the limit $\beta\to\infty$ and rescaling $\hbar$ so that $\gamma \neq 0$, assuming that the limit procedure and the series expansion for the potential commute, we have:
\begin{equation}\label{eq:Haffine}
h(X,P_X)=\frac{1}{4}P_{X}^{2}+e^{-2Q(X)}+\frac{\gamma}{4}X^{-2} \ ,
\end{equation}
where we renamed $\gamma(2)\to\gamma$, as our starting point, with the equations of motion:
\begin{subequations}\label{eq:AHeom}
\begin{align}
P_{X,r} = & -2Q'e^{-2Q}+\frac{\gamma}{2}X^{-3}\ ,\\
X_{r} = & \frac{1}{2}P_{X}
\end{align}
\end{subequations}
and the constraint $h(X,P_X)=0$.
\section{Physics of affine black holes}
\label{s:physics}
\subsection{Thermodynamics}
\label{s:thermo}
It is interesting to see what kind of effect the addition of quantum dynamics can have on the thermodynamical properties of BH solutions in the semiclassical regime. An extremely complete description of BH thermodynamics in the framework of dilaton gravity can be found in \cite{Grumiller:2007ju} and we will base our analysis mainly adapting that procedure to our effective model, referring the interested reader to that paper for further details.
\subsubsection{A summary on black hole thermodynamics}
In \cite{Grumiller:2007ju} the authors consider the full 2dGDT action \eqref{eq:DGaction}, with Euclidean signature, and focus on BH solutions given by \eqref{eq:diaggauge} and \eqref{eq:DGsol}. This induces a periodicity in the time coordinate given by:
\begin{equation}\label{eq:beta}
\bar{\beta}=\left.\frac{4\pi}{w'(X)}\right|_{X_{h}} = T^{-1}\ ,
\end{equation}
where $X_h$ is the value of the dilaton field at the horizon, $T$ the temperature measured by an asymptotic observer and is also related to surface gravity. In order to build a canonical ensemble, motivated the path integral formulation first developed in \cite{York:1986it}, one can introduce a thermal reservoir that would, for instance, fix the value of the dilaton charge $D$ at the boundary. Taking advantage of the simplifications brought in by working in two-dimensions, one can choose $D(X)=X$, and therefore define the thermal reservoir by having an upper bound on the dilaton $X \le X_c$, where the subscript $c$ indicates quantities calculated at the cavity wall. At the same time the thermal reservoir fixes also the value of the local temperature $T_c$, related to the period \eqref{eq:beta} by:
\begin{equation}\label{eq:betac}
\bar{\beta}_c=\bar{\beta} \sqrt{\xi_c} = T_c^{-1}\ .
\end{equation}
This restricts the number of BH solutions included in the ensemble, but it is also possible in certain cases to take the limit $X_c\to\infty$ later on. The partition function for the ensemble is build with a Euclidean path integral:
\begin{equation}
\mathcal{Z}=\int \mathcal{D}g \mathcal{D}X \exp{-\Gamma[g,X]}\ ,
\end{equation}
over all solutions included in the ensemble.\\
In the semiclassical limit the most relevant contribution to $\mathcal{Z}$ is given by the minimum of the action, which corresponds to classical solutions, where first order variations of $\Gamma$ are vanishing. If the on-shell action is finite, assuming that the quadratic variation is positive definite, the path integral can be approximated as:
\begin{equation}
\mathcal{Z}\sim\exp{-\Gamma_c}\times\left(\text{quadratic terms}\right)\ ,
\end{equation}
where $\Gamma_c$ is the on-shell action and it is calculated for solutions that fit in the cavity, i.e. $X_h \leq X \leq X_c$. In order to ensure these conditions are satisfied it is important to include suitable boundary terms \cite{Grumiller:2007ju,Bergamin:2007sm}. It is then possible to proceed in the explicit calculation of thermodynamical quantities:
\begin{subequations}\label{eq:thermodynamics}
\begin{align}
\text{free energy}\quad  & F_{c}=-T_{c}\ln\mathcal{Z}\simeq T_{c}\Gamma_{c}\ ,\\
\text{entropy}\quad  & S=-\frac{\partial F_{c}}{\partial T_{c}}\ ,\\
\text{internal energy}\quad & E_{c}=F_{c}+T_{c}S\ ,\\
\text{specific heat}\quad & C_{c}=-\frac{\partial E_{c}}{\partial T_{c}}=T_{c}\frac{\partial S}{\partial T_{c}}\ .
\end{align}
\end{subequations}
An important condition for the thermodynamic stability of the ensemble is that the ground state, i.e. the state with minimal $\Gamma_c$, is characterized by a positive specific heat $C_c>0$. Condition that has to be maintained in the limit $X_c\to\infty$ if we want to remove the cavity wall.
\subsubsection{Thermodynamics for the effective black hole model}
Let us now turn our attention to the effective BH model described by \eqref{eq:XIsol} and \eqref{eq:Leff}. As we mentioned before, this model is physically equivalent to the full dilaton gravity model, hence we can inherit the procedure described above, with a few caveats.\\
The thermal reservoir is introduced by fixing an upper value to the dilaton $X\le X_c$.\\ The period $\bar{\beta}$, as defined in \eqref{eq:beta}, will have to include possible corrections to the value of the dilaton field at the horizon. By retaining first order corrections in $\gamma$ only, we can write the period for ABH as:
\begin{equation}
\bar{\beta}_{ABH} = \bar{\beta} + \alpha \gamma + O\left(\gamma^2\right) \ ,
\end{equation}
where $\alpha$ is some coefficient. At the cavity wall, on the other hand, quantum corrections can be neglected, since the potential term enforced by $\gamma$ is suppressed.
We can then rewrite \eqref{eq:betac} as:
\begin{equation}\label{eq:betacabh}
\left(\bar{\beta}_{ABH}\right)_c=\bar{\beta}_c + \alpha' \gamma \ ,
\end{equation}
where $\alpha'$ incorporates the Tolman factor $\sqrt{\xi_c}$. As we will see further on, explicit calculation of the factor $\alpha'$ is not necessary, as it will only contribute to second order corrections to the free energy $F_c$.\\
Let us now look at the path integral built with the effective action associated with \eqref{eq:Haffine} for the semiclassical case with $\beta \to \infty$ (a discussion on the case for finite $\beta$ is in Appendix \ref{s:finitebeta}), restricted to the interval $r_h\le r \le r_c$, that takes the form:
\begin{equation}
\Gamma = \int_{r_h}^{r_c} dr\left[\frac{g}{N}X_{r}^{2}+Ne^{-2Q}-\frac{\gamma N}{4g}X^{-2}\right]\ .
\end{equation}
A vanishing first order variation of this action requires the addition of a boundary counterterm $I_{ct}=\int dr\partial_{r}F(X)$. In the gauge $g=N=1$ it is easy to check that we need:
\begin{equation}
F(X)=-2\int^{X}d\tilde{X}\left(e^{-Q(\tilde{X})}-\frac{\gamma}{4}e^{Q(\tilde{X})}\tilde{X}^{-2}\right)\ .
\end{equation}
Since we are looking specifically at the interval $r_h\le r \le r_c$, were $X(r)$ is one-to-one map to $X_h\le X \le X_c$, we can replace integrals in $r$ with integrals in $X$. In order to do so we can look at the constraint equation $h=0$:
\begin{equation}
X_{r}^{2}=e^{-2Q}-\frac{\gamma}{4}X^{-2} 
\end{equation}
and, by completing the square on the r.h.s. and limiting ourselves to first order corrections in $\gamma$, we can obtain:
\begin{equation}\label{eq:firstorderxr}
X_r=e^{-Q}-\frac{\gamma}{8}e^{Q}X^{-2}+O\left(\gamma^2\right)\ ,
\end{equation}
which also solves the semiclassical equations of motion \eqref{eq:AHeom}. This solution can be inverted for a change of variable in the integrals, resulting in the first order approximation in $\gamma$:
\begin{equation}
dr \simeq e^{Q}\left(1-\frac{\gamma}{8}e^{2Q}X^{-2}\right)^{-1}dX \simeq e^{Q}\left(1+\frac{\gamma}{8}e^{2Q}X^{-2}\right)dX\ .
\end{equation}
Finally, we can calculate the improved action $\Gamma+I_{ct}$ on-shell, as an integral in $X$, and we have:
\begin{equation}\label{eq:onshellaction}
\Gamma_c=\frac{\gamma}{4}\int_{X_{h}}^{X_{c}}dX\left[e^{Q}X^{-2}\right]+O\left(\gamma^{2}\right)\ ,
\end{equation}
where only $\gamma$ corrections have survived. It is now clear that if $F_c=T_c\Gamma_c$ the first order correction to $\bar{\beta}_c=T^{-1}_c$ in \eqref{eq:betacabh} can be neglected.\\
In order to calculate the thermodynamical quantities \eqref{eq:thermodynamics}, we can use the definition of $T_c$ in \eqref{eq:betac}, at fixed $X_c$, to turn derivatives w.r.t. $T_c$ in derivatives w.r.t. $X_c$, as in:
\begin{equation}
\frac{\partial}{\partial T_{c}}=\left( \frac{\partial T_c}{\partial X_c} \right)^{-1}\frac{\partial}{\partial X_{c}}\ .
\end{equation}
We can now turn our attention to some of the most important BH models in dilaton gravity by choosing the specific form of the potentials $Q,w$, looking at how the quantum corrections in $\gamma$ (might) affect thermodynamical properties. 
Let us stress that by looking at the effective action for $X$, rather than the full action \eqref{eq:DGaction}, we are building a fundamentally different ensemble which limits the usefulness of our construction to specific tasks.\\
Clearly, with no classical contributions in \eqref{eq:onshellaction} and no information on BH solutions nor geometry, we cannot ensure straightforwardly the validity of relations such as the area law $S \sim A$ or the equivalence of quasi-local energy \cite{Brown:1992br} with internal energy, which would require further investigation, going beyond the scope of this work. We will see however that an agreement between the sign of entropy corrections and the displacement of BH horizons is present in the model.\\
Furthermore, we cannot expect to reproduce expressions known from \cite{Grumiller:2007ju} for thermodynamical quantities, while a certain similarity in general behaviour can be obtained because of the identical expressions for the temperature $T_c$ and the metric component $\xi$. What we can do, however, is look at general thermodynamical features in order to determine whether the inclusion of quantum corrections due to affine quantization changes the physical properties of known BH solutions. In particular we will check if any restriction to the free parameters of the models is required to ensure thermodynamical stability, or if any region in parameter space that is excluded classically might become accessible in the semiclassical regime.
\subsection{Comparing classical and semiclassical solutions}
\label{s:comparing}
While classical solutions are known, the full equations of motion \eqref{eq:AHeom}, in most cases, are not easy to handle analytically. Additionally, first order approximations such as \eqref{eq:firstorderxr} will fail to provide the full dynamics close to the singularity. Therefore we can turn to numerical solutions for the semiclassical regime. In the following we will apply an identical procedure to a number of models of dilaton gravity, comparing classical and semiclassical solutions with identical initial conditions in the asymptotic region, where the effect of the $\gamma$ corrections is negligible. A modified procedure dealing with the case of finite $\beta$ is in Appendix \ref{s:finitebeta}. In particular we:
\begin{enumerate}
\item Calculate analytically the classical solutions $X(r)$, $P_X(r)$.
\item Fix the free parameters to suitable values.
\item Calculate initial conditions $X(r_0)$, $P_X(r_0)$ at $r_0 \gg 0$.
\item Numerically solve the semiclassical equations of motion \eqref{eq:AHeom} in $r\in[-r_0,r_0]$, using the initial conditions calculate above.
\item Check that the constraint $h=0$ is enforced.
\end{enumerate}
The different parameters, e.g. the mass $M$ and the value of the constant $\gamma\sim\hbar^2$ are chosen to best serve in the numerical computations involved. In any case the constraint $h$ is enforced with a numerical tolerance corresponding to $|h|\le 10^{-6}$.\\
In the comparison between the classical and semiclassical regimes we will label quantities with subscripts $(C)$ and $(A)$ respectively, when needed. 
\section{Thermodynamics, singularity removal and horizon corrections}
\label{s:models}
Let us now apply what developed in the previous Section to a number of models in dilaton gravity.
\subsection{The \emph{ab}-family}
\begin{figure}
\includegraphics[width=.7\textwidth]{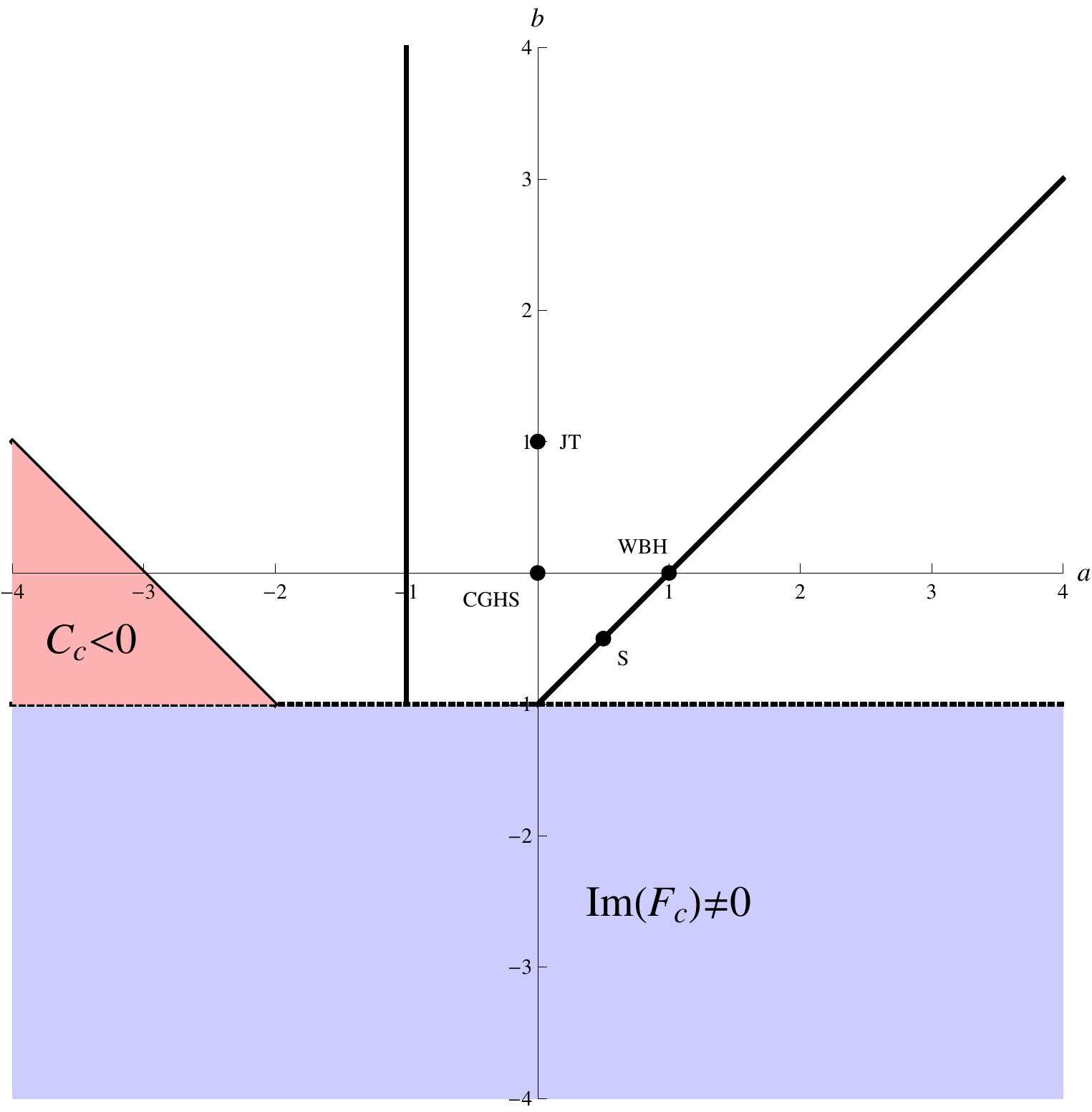}
\caption{Regions of the $ab$ parameter plane with regions of thermodynamical instability and non-physical solutions.}
\label{f:abfamily}
\end{figure}

The {ab}-family encompasses a number of renown BH models, as the dimensional reduction of spherically symmetric BH in $d+1$ dimensions, the Witten BH, CGHS and many more (see \cite{Grumiller:2006rc} and references therein). It is characterized by the potentials:
\begin{equation}
Q = -a\ln X \qquad w = \frac{B}{(b+1)}X^{b+1} \ ,
\end{equation}
which give us an extended Hamiltonian in the form:
\begin{equation}\label{eq:Haffineab}
h(X,P_X)=\frac{1}{4}P_{X}^{2}+X^{2a}+\frac{\gamma}{4}X^{-2} \ .
\end{equation}
The on-shell improved action \eqref{eq:onshellaction} that reads:
\begin{equation}
\Gamma_{c}=\frac{\gamma\left(X_{h}^{-a-1}-X_{c}^{-a-1}\right)}{4(a+1)}\ ,
\end{equation}
while the temperature at the cavity wall is given by:
\begin{equation}
T_{c}=\frac{X_{h}^{b}\sqrt{B\left(b+1\right)}X_{c}^{\frac{a}{2}}}{4\pi\sqrt{\left(X_{c}^{b+1}-X_{h}^{b+1}\right)}}\ .
\end{equation}
As we are interested in removing the bound on $X$ and take the limit $X_c\to\infty$, we need to make sure that $T_c$ stays real and non-negative for all values of $X_c$, which is the case if the sign of $B(b+1)$ is kept positive. We can therefore calculate all thermodynamical quantities \eqref{eq:thermodynamics} and look at their behaviour while $X_c\to\infty$, keeping an eye on the specific heat. Assuming $B=b+1$ for simplicity, with no loss of generality, we can divide the $ab$-plane is a number of regions with different asymptotic behaviour, summarized in Table \ref{t:abfamily} and visualized in Figure \ref{f:abfamily}.
\begin{table}
\begin{tabular}{|p{.4cm} C{4.5cm}|C{1.5cm}|C{1.5cm}|C{1.5cm}|C{1.5cm}|C{1.5cm}|}
\hline
 & & $T_c$ & $F_c$ & $S$ & $E_c$ & $C_c$\\
\hline
 1) & $a>b+1 \land b>-1$ & $\infty$ & $\infty$& $S<0$& $0$ & $0^+$ \\
 2) & $a=b+1 \land b>-1$ & $T_c>0$ & $F_c>0$ & $S<0$ & $0$ & $0^+$ \\
 3) & $-1<a<b+1 \land b>-1$ & $0$ & $0$ & $S<0$ & $0$ & $0^+$ \\
 4) & $a=-1 \land b>-1$ & $0$ & $0$ &$-\infty$ & $0$ & $C_c>0$ \\
 5) & $-b-3<a<-1 \land b>-1$ & $0$ & $0$ & $-\infty$ & $0$ & $\infty$ \\
 6) & $a=-b-3 \land b>-1$ & $0$ & $F_c>0$ & $S>0$ & $0$ & $\infty$ \\
 7) & $a<-b-3 \land b>-1$ & $0$ & $\infty$ & $\infty$ & $\infty$ & $-\infty$\\
 \hline
  & $b<=-1$ & \multicolumn{5}{c|}{Excluded by $\text{Im}(F_c)\neq 0$ and $\text{Im}(E_c)\neq 0$}\\
 \hline
\end{tabular}
\caption{Summary of asymptotic $X_c\to\infty$ values for thermodynamical quantities for the \emph{ab}-family, for different regions on the $ab$-plane.}\label{t:abfamily}
\end{table}
The region $b\le -1$ is not physical, being characterized by an imaginary value for the free energy $F_c$ and/or the internal energy $E_c$, which can only be avoided by taking $B(b+1)<0$, contradicting the condition required for the positivity of the temperature at the cavity wall. The same region is also excluded in \cite{Grumiller:2007ju}, as a result of thermodynamical instability. The only additional excluded region w.r.t. \cite{Grumiller:2007ju} is $a<-b-3 \land b>-1$, characterized by a negative specific heat, thus resulting in thermodynamical instability. This is the only significant difference, while the rest of the plane, in particular where renown models are located, is characterized by real and (mostly) finite thermodynamical quantities, so that including affine quantization effects is generally compatible with known results from standard 2dGDT. In particular the specific heat $C_c$ is always non-negative, approaching zero only from above in the limit $X_c\to \infty$.\\
Notice also that all entropy corrections are negative, with the exception of the line $a=-b-3 \land b>-1$, boundary of the excluded region with negative specific heat. We will see later on, when comparing classical and semiclassical solutions, that there seems to be a correspondence between the sign of these corrections and the displacement of BH horizons.\\
Let us now look, as a way of example, at a comparison between classical and semiclassical solutions for a number of models within the \emph{ab}-family. We expect the $\gamma$ corrections to induce the removal of singularities, at least for all cases with $a>-1$, due to the form of the potentials in \eqref{eq:Haffineab}.
\subsubsection{Schwarzschild BH}
\begin{figure}
\includegraphics[width=.9\textwidth]{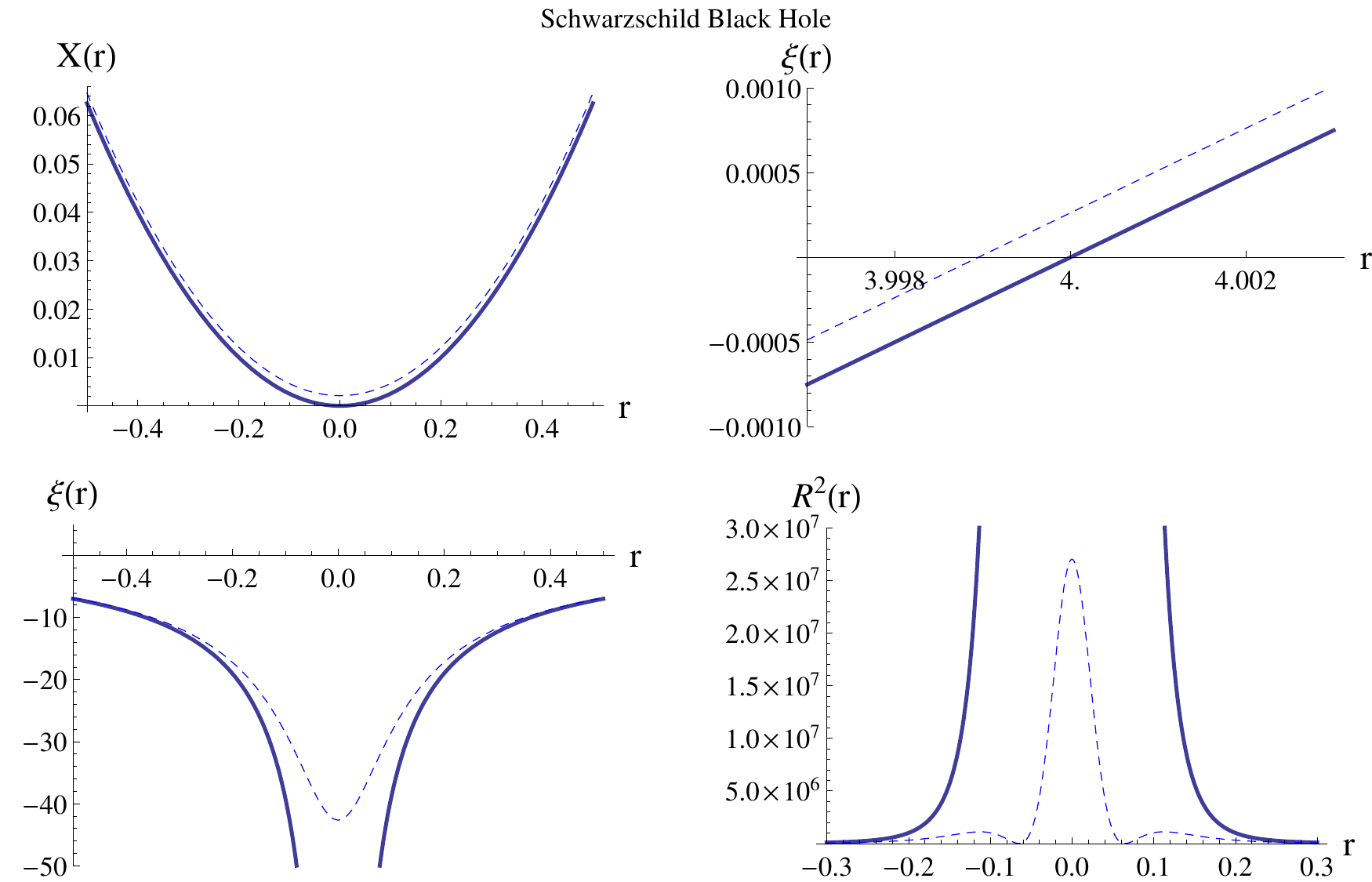}
\caption{Comparison of classical (continuous line) and semiclassical (dashed line) solutions for the Schwarzschild BH model, for parameters $\gamma=10^{-10}$, $M=1$, $B=1/2$, $r_0=10^3$. }
\label{f:schwarz}
\end{figure}

The Schwarzschild BH (see Figure \ref{f:schwarz}), obtained as a dimensional reduction of $3+1$ dimensional spherically symmetric BH solutions, is characterized by the parameters $a=1/2$, $b=-1/2$, with $B>0$ (cf. Figure \ref{f:abfamily}). Classical solutions are defined on $r\in \mathbb{R}$, with a curvature singularity in $r=0$ and horizons in $r=\pm r_{h(C)}$, while $\xi\to - \infty$ when $r\to0$.\\
In the semiclassical regime the presence of $\gamma$ corrections modifies the behaviour of $X$ close to the origin, avoiding the reaching of $X=0$, and determines both the removal of the curvature singularity and a smaller extension of the black hole itself. The metric component $\xi$ is finite at $r=0$, as is the Kretschmann scalar $K=R^2$. The local curvature is still quite large close to the origin, with a ratio between $K$'s at $r=0$ and at the horizon greater than $10^9$.\\
The location of the horizon is also modified, with a normalized difference $r_{C/A}=\frac{r_{h(A)}-r_{h(C)}}{r_{h(C)}}\sim -10^{-4}$, which is in agreement with the sign of the first order entropy calculation, for which $S_{(A)}<0$ while $S_{(C)}=0$.
\subsubsection{Jackiw-Teitelboim BH}
\begin{figure}
\includegraphics[width=.9\textwidth]{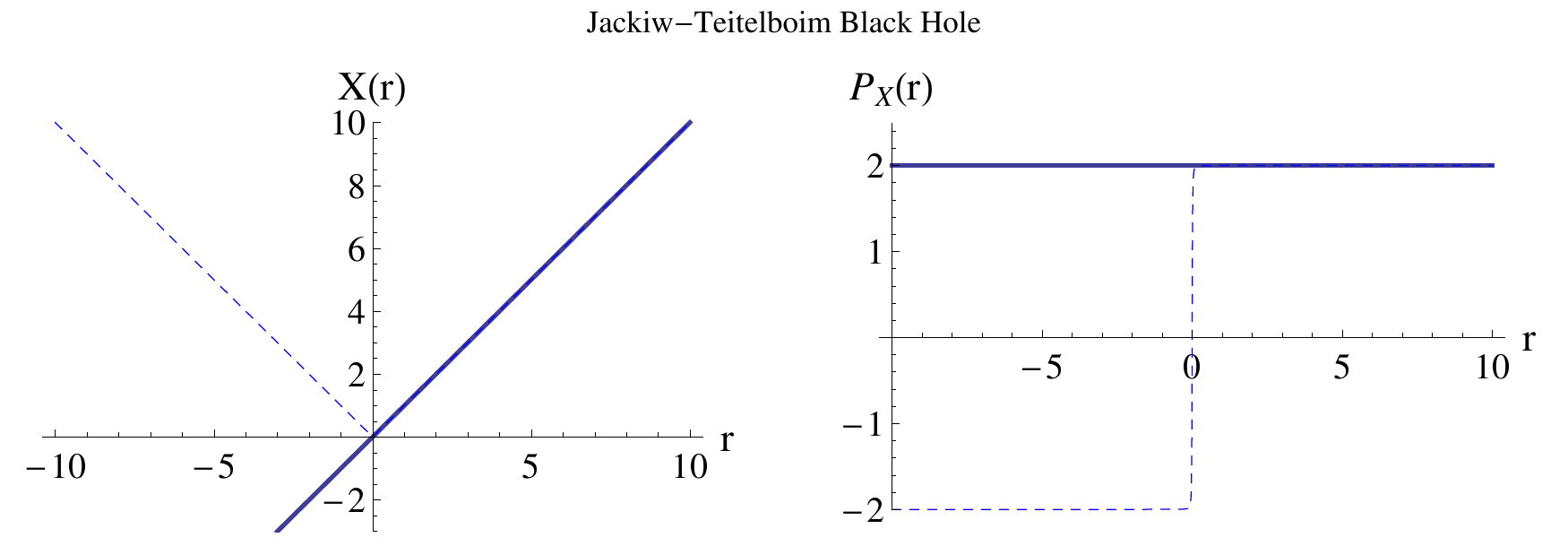}
\caption{Comparison of classical (continuous line) and semiclassical (dashed line) solutions for the Jackiw-Teitelboim BH model, for parameters $\gamma=10^{-4}$, $M=40$, $B=1$, $r_0=10^3$. }
\label{f:JT}
\end{figure}

The Jackiw-Teitelboim model (Figure \ref{f:JT}) corresponds to $a=0$, $b=1$ and its classical solutions present an horizon at $r=r_{h(C)}$, but no curvature singularity. Quantum corrections induce a rebound of $X(r)$ close to $r=0$, so that for negative $r$ we now have $X>0$. The horizon is also shrunk with $r_{C/A}\sim- 10^{-6}$ with our choice of parameters, again in agreement with entropy in the semiclassical regime.
\subsubsection{Witten BH}
\begin{figure}
\includegraphics[width=.9\textwidth]{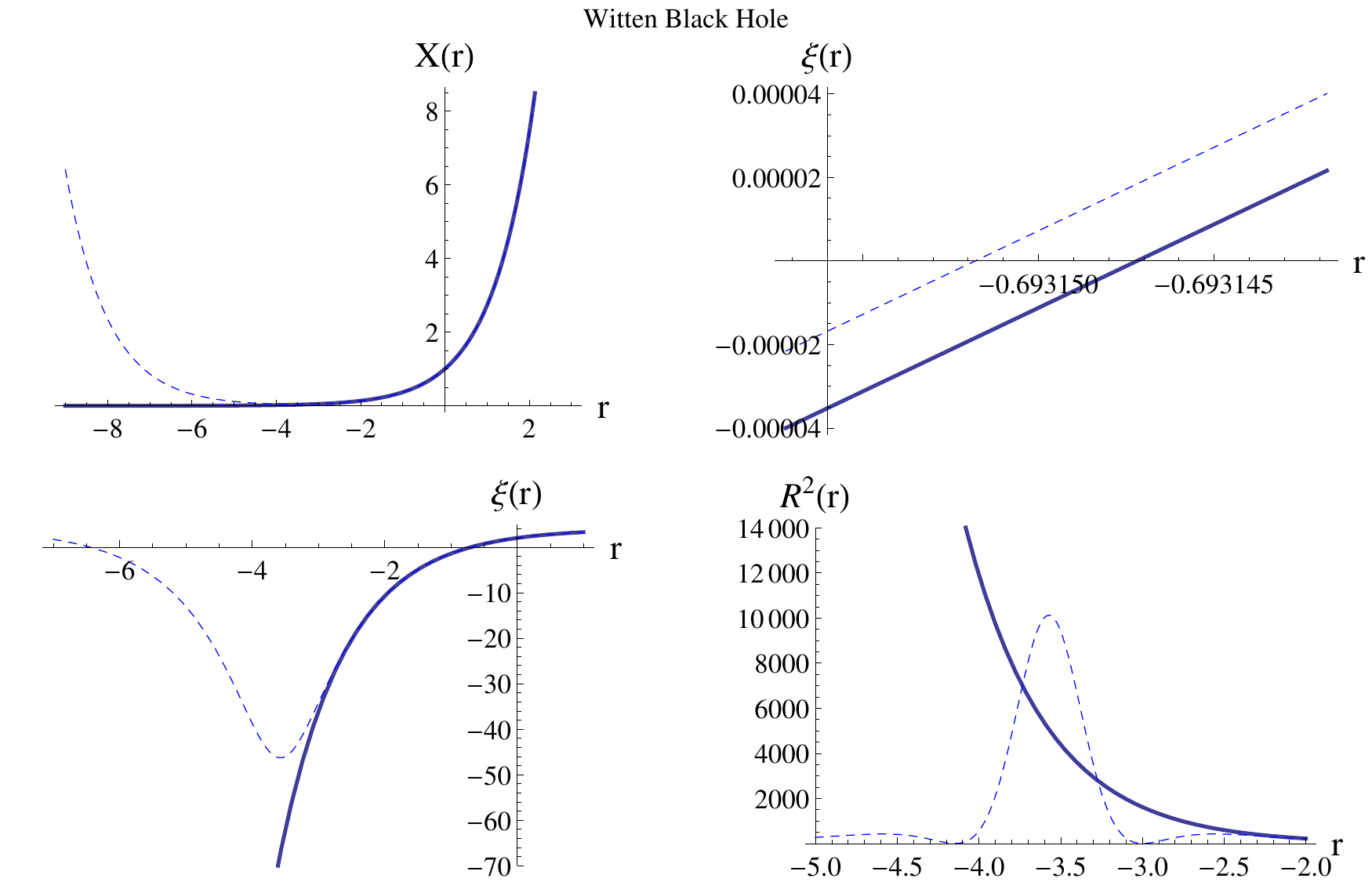}
\caption{Comparison of classical (continuous line) and semiclassical (dashed line) solutions for the Witten BH model, for parameters $\gamma=10^{-5}$, $M=1$, $B=4$, $r_0=10$. }
\label{f:WBH}
\end{figure}

The Witten BH model (Figure \ref{f:WBH}) is obtained with $a=1$, $b=0$. Classical solutions consist in $X\sim e^r$, with an horizon at a single $r$-value and a curvature singularity located at $r\to -\infty$, where also $\xi\to -\infty$. While classically $X$ goes to $0$ with $r\ll0$, quantum corrections induce a bounce of $X(r)$ for negative $r$. This corresponds to a finite value for both $\xi$ and $K$, so that the singularity is effectively removed. The horizon is also modified, agreeing with the sign of entropy, with again $r_{C/A}\sim -10^{-6}$ and a second horizon appearing on the left side of the new minimum of $\xi$.
\subsubsection{Other models of the \emph{ab-family}}
The same analysis can be carried out, with the appropriate choice of parameters, for any point on the \emph{ab}-plane. It leads to similar results in all the cases with $a>-1$, including the CGHS model ($a=b=0$), which exhibits a bounce similar to what happens in the Jackiw-Teitelboim case, with also $r_{C/A}<0$.\\
When $a<-1$ the singularity is indeed not removed, but shifted to smaller values of $r$. In these cases the singularity is usually due to the presence of a $2n$th-root $r^{1/2n}$, so that the effect of the $\gamma$ corrections is only moving the value of $r$ where the root becomes imaginary. When horizons are present, however, they are also moved, in agreement with the sign of entropy corrections.
\subsection{Liouville Gravity}
\begin{figure}
\includegraphics[width=.9\textwidth]{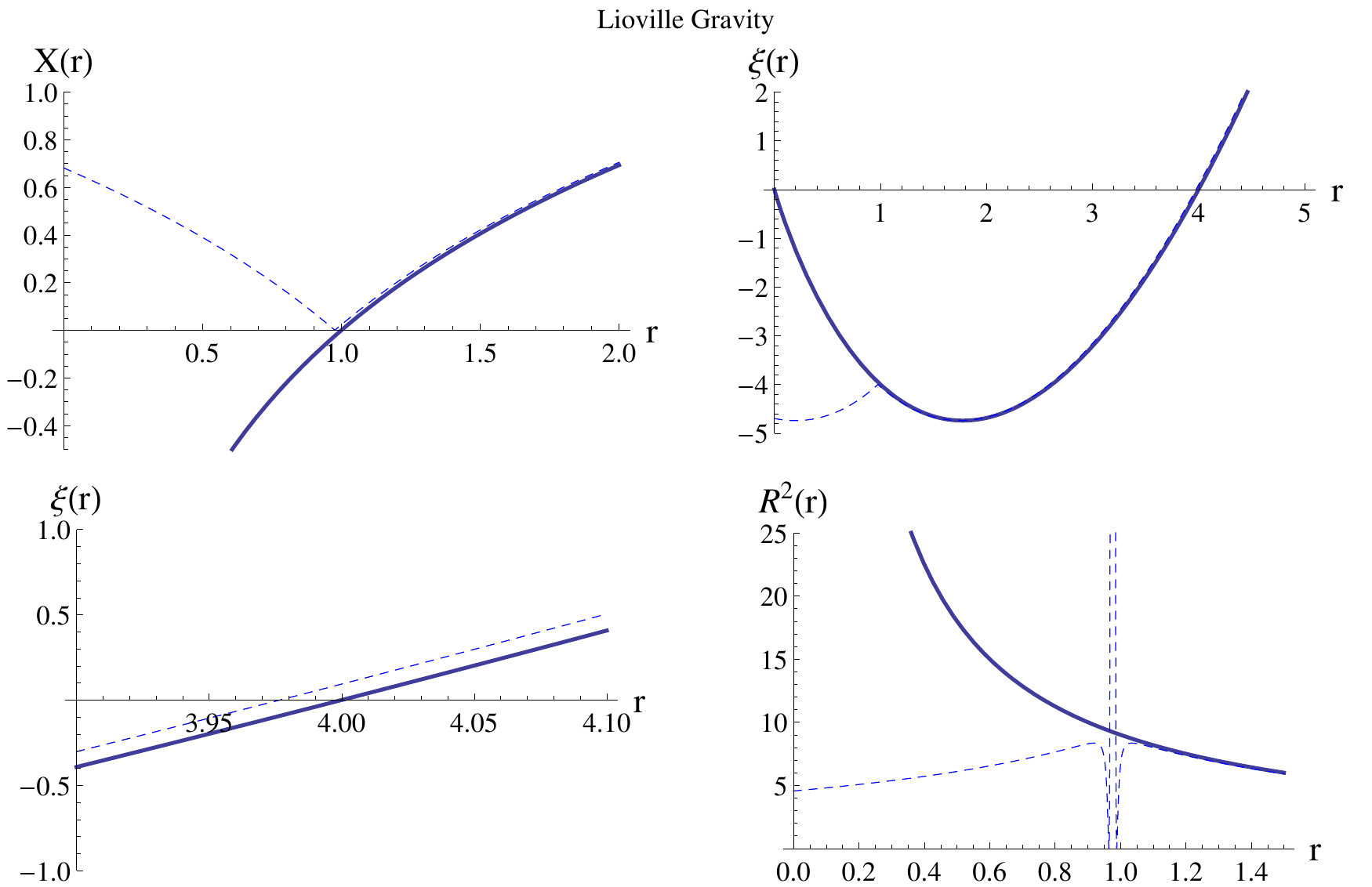}
\caption{Comparison of classical (continuous line) and semiclassical (dashed line) solutions for Liouville Gravity, for parameters $\gamma=10^{-5}$, $M=4$, $a=1$, $\alpha=-1/2$, $b=-1$, $r_0=100$. }
\label{f:liouville}
\end{figure}

Liouville Gravity is obtained with the potentials:
\begin{equation}
Q=aX \qquad w(X)=-\frac{2b}{a+\alpha} \exp{(a+\alpha) X} \ .
\end{equation}
The temperature at the cavity wall is then given by:
\begin{equation}
T_{c}=-\frac{be^{(a+\alpha)X_{h}}}{2\pi\sqrt{e^{aX_{c}}\left(-\frac{2be^{(a+\alpha)X_{c}}}{a+\alpha}-2M\right)}} \,
\end{equation}
so that in order for $T_{c}$ to be real and non-negative, we need to have:
\begin{eqnarray*}
b & < & 0\ ,\\
a+\alpha & > & 0\ .
\end{eqnarray*}
Solutions to the classical e.o.m. are given by two branches $X(r)=a^{-1}\ln \left(\pm a r \right)$, with the appropriate choice of sign to ensure the existence of the log. In the limit of large $X_c$ the leading contribution to the specific heat is given by:
\begin{equation}
C_c \sim \frac{\gamma \left(\alpha X_{c}\right)}{2\left(2a+\alpha\right)^{2}} e^{aX_{c}}X^{-2}_c \,
\end{equation}
so that to ensure a non-negative limit we have to exclude the region $\alpha<0 \land a>0$, which reduces the allowed region on the $a\alpha$-plane to $(\alpha \leq 0\land a>-\alpha )\lor (\alpha >0\land a>0)$. This additional restriction is not required in the case of full 2dGDT. We are then allowed to remove the cavity wall to infinity and calculate the other limits:
\begin{eqnarray*}
\alpha >0\land a>0 & : & \qquad F_c\to 0^+\qquad S\to -\infty \qquad E_c \to 0^+ \ ,\\
\alpha \leq 0\land a>-\alpha & : & \qquad F_c\to \infty \qquad S\to -\infty \qquad E_c \to \infty \ .
\end{eqnarray*}
Let us now compare classical and semiclassical solutions (see Figure \ref{f:liouville}). As mentioned above classical solutions are made of two distinct branches, depending on the sign of $r$, and they are overall symmetric w.r.t. the $r=0$ axis. The dilaton $X$ diverges to $-\infty$ when $r\to0$, with a corresponding curvature singularity for $\alpha<0$ and a single horizon at which $\xi(r_h)=0$. As in the case of the \emph{ab}-family, the $\gamma$ corrections induce a bounce in $X$, which is kept positive, and the removal of the curvature singularity, so that $K>0$ at $r=0$. At the location of the bounce a sharp, but finite, curvature peak is present and, compatibly with the sign of the entropy correction, the horizon has a smaller extent. Similar results are obtained for $\alpha>0$, case in which there is no classical singularity to be removed.
\subsection{Schwarzschild-(A)dS}
\begin{figure}
\includegraphics[width=.9\textwidth]{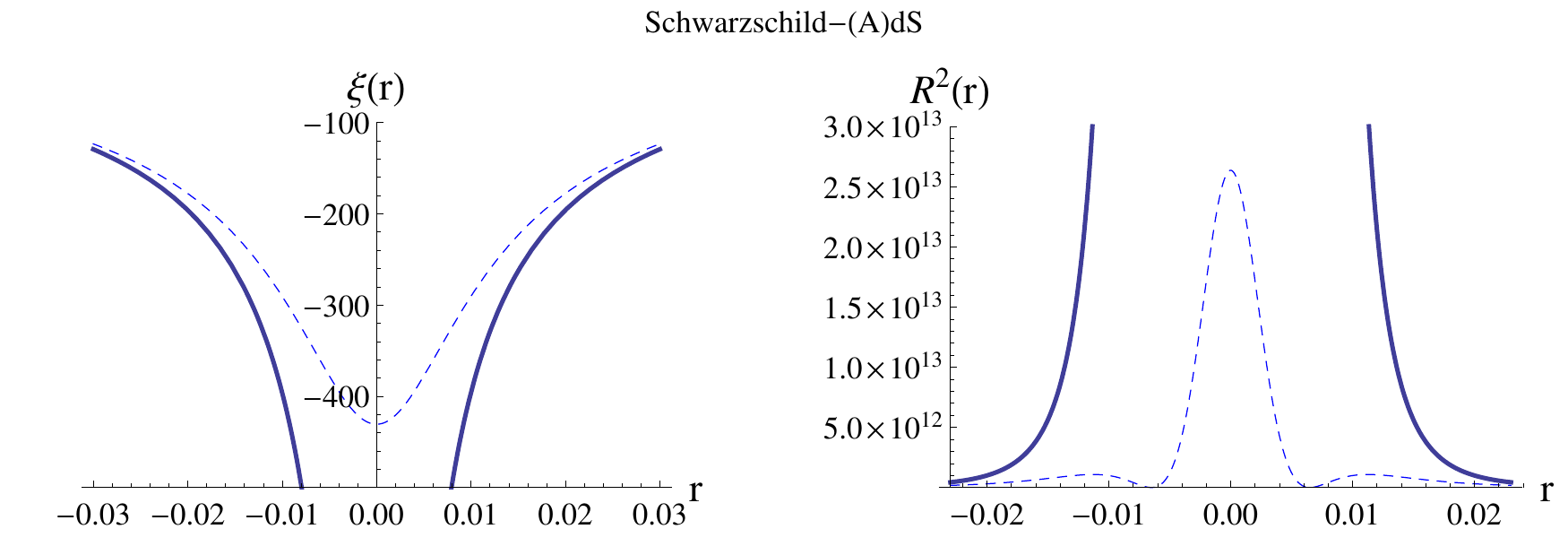}
\caption{Comparison of classical (continuous line) and semiclassical (dashed line) solutions for the Schwarzschild-(A)dS model, for parameters $\ell = 1$, $G_4=1/2$ $M=1$ and $\gamma = 10^{-10}$. }
\label{f:sads}
\end{figure}

Let us consider one last model in our analysis. The Schwarzschild-(A)dS BH is obtained by fixing the potentials to:
\begin{equation}
Q=\sqrt{\frac{G_4}{2X}}\qquad w=\sqrt{\frac{2X}{G_4}}\left(1+\frac{2G_4}{\ell^2}X\right)\ ,
\end{equation}
which allow us to look at an extended Hamiltonian only in the $\beta\to\infty$ limit. The temperature is positive definite for all values of the parameters:
\begin{equation}
T_{c}=\frac{6G_{4}X_{h}+\ell^{2}}{4\ 2^{3/4}\pi G_{4}^{1/4}\ell^{2}X_{h}}e^{-\sqrt{\frac{G_{4}}{2X_{c}}}}\left(X_{c}^{1/2}-X_{h}^{1/2}+2G_{4}\left(X_{c}^{3/2}-X_{h}^{3/2}\right)\right)^{-1/2}>0\ .
\end{equation}
As the explicit expressions for \eqref{eq:thermodynamics} are rather complicated, we will look, as a way of example, at a numerical computation of the limits $X_c \to \infty$. For instance, with a parameter choice of $\ell = 1$, $G_4=1/2$, $M=1$ and $\gamma = 10^{-10}$ we obtain:
\begin{equation}
F_c\to0^+ \qquad S<0 \qquad E_c \to 0^+ \qquad C_c \to 0^+\ .
\end{equation}
Classical solutions, with these parameters, exhibit a curvature singularity at $r=0$, again removed in the semiclassical regime (see Figure \ref{f:sads}), alongside a reduced size for the BH, with $r_{C/A}\sim - 10^{-6}$.
\section{Discussion}
\label{s:discussion}
In this work we have developed an extensive analysis of an effective description of black holes in the semiclassical limit of affine coherent state quantization given by the Weak Correspondence Principle.\\
Starting from the general 2dGDT action, in the diagonal gauge typical for static black holes solutions, we have built an effective model where the space-time metric component $\xi$ is taken to be the known classical solution to the 2dGDT equations of motion. This had the purpose of leaving the notion of classical geometry of space-time untouched by direct quantum corrections, which only appeared through the dynamics of the remaining field, the dilaton $X$. An appropriate action, inclusive of one-dimensional diffeomorphisms invariance, was built from the equations of motion. The reduction to a one-dimensional one-field system allowed us to use the affine coherent state quantization scheme in order to obtain a semiclassical model, inclusive of quantum corrections proportional to $\hbar^2$.\\
Our main goals were to investigate the effect of these quantum corrections on both the thermodynamical properties of know models of dilaton gravity and the dynamics of their solutions.\smallbreak
An interesting question was to see whether the semiclassical regime would be compatible with known results from dilaton gravity (for instance inducing thermodynamical instabilities or imaginary limits) and whether the requirement of physicality would restrict free parameters.\\
BH thermodynamics has been studied in the canonical ensemble approach developed for full 2dGDT, adapted to using our effective action in building the partition function. A comparison with known results shows a large degree of compatibility: excluded regions in the classical regime are identically excluded in the semiclassical case, while only minor additional restrictions on the free parameters are introduced in order to ensure thermodynamical stability. In any case no region excluded classically becomes accessible in the semiclassical limit.\\
In particular the \emph{ab}-family (see Figure \ref{f:abfamily}), besides the known excluded region $b<-1$, requires also $a>-b-3$, condition that however does not affect any of the renown models (such as CGHS, Witten, Jackiw-Teitelboim, Schwarzschild BHs). Also Liouville Gravity requires a stronger restriction on one of its parameters, in any case still allowing for both singular and non-singular solutions. In the case of Schwarzschild-(A)dS no restriction is required on the free parameters (which in this instance are Newton's constant and the cosmological constant).\\
This overall compatibility is non trivial, considering the profound differences in the construction of the ensemble, and could be seen as a positive sign on the viability of affine coherent state quantization.\smallbreak
For a number of different parameter choices, classical and semiclassical solutions have been compared, using numerical methods for the latter. The introduction of quantum corrections is indeed able to eliminate curvature singularities, which are replaced by large but finite values for the Kretschmann scalar $K=R^2$. Only exception are those models in the \emph{ab}-family that are plagued by high order $X^{-n}$ singularities. It is not to be excluded, however, that non-perturbative effects of the full affine coherent state path integral might dispose of those singularities as well.\\
In addition, all semiclassical solutions exhibit a displaced horizon w.r.t. corresponding classical solutions with identical initial conditions. The sign of this displacement is compatible with the sign of the entropy corrections calculated in the canonical ensemble.\smallbreak
A clear limit of our analysis lies in the fact that we have taken into account a very specific effective model in a semiclassical limit and it is of course of interest, in the light of the positive results above, to relax these constraints.\\
We have focused on the case $\beta \to \infty$, mainly motivated by the simplified picture in which only an additional potential term appears in the semiclassical Hamiltonian and multiplicative factors to the classical terms are reduced to unity. While the comparison between classical and semiclassical solutions can be easily repeated for finite $\beta$, leading to the same general properties for the dynamics (singularity removal and modified horizons), it would be interesting to further investigate the role and possible physical relevance of the parameter $\beta$.\\
Furthermore we have allowed no quantum dynamics to enter directly in the metric component $\xi$, relegating a dynamical variable to a fixed function of $X$. Obtaining a semiclassical extension of the full 2dGDT action, instead of the effective model we used, would allow us to carry on a more complete analysis of both the thermodynamics and the solutions. A principal difficulty to this end is given by the highly non-linear couplings between $\xi$ and $X$, which would complicate the calculation of an extended Hamiltonian with the methods used in this work. A simple workaround would be the restriction to those 2dGDTs models that are dual to two decoupled Liouville fields \cite{Zonetti:2011ky}.\\
Comparison with other quantization approaches, e.g. Loop quantum gravity, Higher Spin Gravity, polymer quantization, would prove of sure interest. More formal extensions would consider alternative choices for the fiducial vector in the quantization procedure, as well as the application of the coherent state path integral quantization on the 2dGDT action.
\appendix
\section{Finite $\beta$} \label{s:finitebeta}
If we keep $\beta$ finite in \eqref{eq:Haffinedelta}, the $\delta$ factors need to be carried on throughout our analysis, with the corresponding lower bounds on the value of $\beta$. In particular, starting from \eqref{eq:Haffinedelta}, we can build the Lagrangian:
\begin{equation}
\mathcal{L}=\frac{gX_{r}^{2}}{N\delta(2)}+N\breve{Q}^{2}(X)-\frac{N\gamma}{g4}X^{-2} \ ,
\end{equation}
where we renamed the corrected potential $\langle e^{-Q(X)}\rangle=\sum Q_n X^n \delta(-n) \to \breve{Q}$ and $\gamma(2)\to\gamma$. $\beta$ is for now limited to $\beta>1$ in order to keep $\delta(2)$ finite. Classical solutions for $X$, to the first order in $\gamma$, are given by:
\begin{equation}
X_{r}=\sqrt{\delta(2)}\left(\breve{Q}-\frac{\gamma}{8}\breve{Q}^{-1}X^{-2}\right)+O\left(\gamma^{2}\right)\ .
\end{equation}
Denoting with $\breve{}$ quantities calculated for finite $\beta$ and including the appropriate boundary counterterm the improved action takes then the form:
\begin{equation}
\breve{\Gamma}=\int dr\left[\frac{gX_{r}^{2}}{N\delta(2)}+N\breve{Q}^{2}(X)-\frac{N\gamma}{g4}X^{-2}\right]-\frac{2}{\sqrt{\delta(2)}}\int^{X_c}_{X_h} d{X}\left(\breve{Q}({X})-\frac{\gamma}{4}\breve{Q}^{-1}{X}^{-2}\right) \ ,
\end{equation}
which on-shell is:
\begin{equation}
\breve{\Gamma}_{c}=\frac{\gamma}{4\sqrt{\delta(2)}}\int^{X_c}_{X_h}\! \! dX\left[\breve{Q}^{-1}X^{-2}\right] \ .
\end{equation}
Thermodynamical quantities \eqref{eq:thermodynamics} are then obtained as before, and will include $\delta$ factors coming from $\breve{\Gamma}_c$, while the temperature $T_c$ does not depend on $\beta$. Fixing the form of the potentials and calculating explicitly $\breve{Q}(X)$ we can look at how a finite value for $\beta$ affects the thermodynamics results of section \ref{s:thermo}.\\
For the \emph{ab}-family we have simply $\breve{Q}=\delta(-a)X^{a}$, which gives:
\begin{equation}
\breve{\Gamma}_{c}=\frac{\gamma\left(X_{h}^{-a-1}-X_{c}^{-a-1}\right)}{4\sqrt{\delta(2)}(a+1)\delta(-a)} = \frac{\bar{\Gamma}_c}{\sqrt{\delta(2)}\delta(-a)}  \ ,
\end{equation}
with a bound $\beta>-a/2$, leading to thermodynamical quantities which are the same as Section \ref{s:thermo} times a factor $\left(\sqrt{\delta(2)}\delta(-a)\right)^{-1}$, hence reproducing all results as far as parameter space is concerned.\\
In Liouville gravity, with its classical potential $Q(X)=aX$, an infinite number of $\delta$ corrections is generated in the calculation of the extended Hamiltonian, which can be resummed and result in:
\begin{equation}
\breve{Q}=\left(\frac{aX}{2\beta}+1\right)^{-2\beta} \ ,
\end{equation}
with no further restriction on $\beta$ than $\beta>1$. The on-shell action will read:
\begin{equation}
\breve{\Gamma}_{c}=\left.-\frac{\gamma\Gamma(1-2\beta)}{4a\sqrt{\delta(2)}}\left(aX+2\beta\right)X^{-2}\left(\frac{aX}{2\beta}+1\right)^{2\beta}\,_{2}\tilde{F}_{1}\left(1,2;2-2\beta;-\frac{2\beta}{aX}\right)\right|_{X_{h}}^{X_{c}}
 \ ,
\end{equation}
where $\,_{2}\tilde{F}_{1}$ is the regularized Hypergeomtric function. Looking at the limit for large $X_c$, the leading contribution to the specific heat takes the form:
\begin{equation}
C_c\sim\frac{-\alpha b^{3}\gamma}{2\left(2a+\alpha\right)^{2}X_{c}^{2}\sqrt{\delta(2)}}\left(\frac{aX_{c}}{2\beta}\right)^{2\beta} \ ,
\end{equation}
excluding the region $a>0\wedge\alpha<0$ as in the case for $\beta\to\infty$. For the free energy $F_c$ we have:
\begin{equation}
F_c \sim \frac{b\gamma(1-2\beta)e^{(a+\alpha)X_{h}}}{8\pi\sqrt{-\frac{2b\delta(2)}{a+\alpha}}X_{c}}\left(\frac{aX_{c}}{2\beta}\right)^{2\beta}e^{-\frac{1}{2}(2a+\alpha)X_{c}} \ ,
\end{equation}
which is non-negative. In the same way, for the entropy:
\begin{equation}
S \sim \frac{\gamma\left(\frac{aX_{c}}{2\beta}\right){}^{2\beta}}{4\sqrt{\delta(2)}(1-2\beta)X_{c}} < 0\ ,
\end{equation}
and for the internal energy:
\begin{equation}
E_c \sim \frac{\gamma\sqrt{(a+\alpha)\left(-b\right)}}{4\sqrt{2}\pi\sqrt{\delta(2)}\left(2a+\alpha\right)}\frac{e^{(a+\alpha)X_{h}}}{X_{c}^{2}e^{(2a+\frac{1}{2}\alpha)X_{c}}}\left(\frac{aX_{c}}{2\beta}\right)^{2\beta} > 0 \ ,
\end{equation}
so that the parameter restrictions for finite $\beta$ coincide with the ones obtained in the $\beta\to\infty$ limit. As mentioned earlier the case of Schwarzschild-(A)dS is only defined for $\beta\to\infty$ due to the form of its $Q(X)$ potential.\\
As for the comparison between classical and semiclassical regimes, we need to modify the procedure described in Section \ref{s:comparing} in order to determine an optimal value of $\beta$ for the numerical solution of the equations of motion. In particular for $\beta\to\infty$ the $\gamma$ correction is negligible at $r_0\gg0$, where initial conditions are calculated, so that the initial values $X_0=X(r_0)$, $P_0=P_X(r_0)$ are solutions to both the classical and extended Hamiltonian constraints and can be used in both cases. For finite $\beta$, on the other hand, because of the $\delta$'s, this will not be true and we will have to find appropriate initial conditions for the semiclassical case depending on the value of $\beta$. We can then follow the modified procedure:
\begin{enumerate}
\item Calculate analytically the classical solutions $X(r)$, $P_X(r)$.
\item Fix all the free parameters (but $\beta$) to suitable values.
\item Calculate initial conditions $X_0=X(r_0)$, $P_0=P_X(r_0)$ at $r_0 \gg 0$ for classical solutions.
\item Solve the constraint $h(P_X,X=X_0,\beta)=0$ for $P_X$, obtaining $P_X = \bar{P}(\beta)$.
\item Find the value of $\beta$ for which $\bar{P}(\beta_0)\sim P_0$, thus matching in the best possible way initial conditions for the classical and semiclassical case.
\item Numerically solve the semiclassical equations of motion \eqref{eq:AHeom} in $r\in[-r_0,r_0]$, using the initial conditions $X_0=X(r_0)$ and $\bar{P}(\beta_0)$.
\item Check that the constraint $h=0$ is enforced.
\end{enumerate}
In this way it is possible to obtain the same singularity avoidance described in Section \ref{s:models}. It is worth mentioning however that any considerations about the properties of the solutions away from either the singularity or the asymptotic region $r_0\gg0$ need to carefully account for the presence of the $\delta$ corrections. These modify the behaviour of the solutions by altering the ratio between kinetic and potential terms in the differential equations, so that features like horizon displacement are no longer obvious in the interplay between genuine $\hbar$-dependent quantum corrections and representation related corrections.
\bibliographystyle{unsrt}
\bibliography{references}
\end{document}